\documentstyle{ioplppt}
\newcommand{\SS}{{\bf S}}
\newcommand{\DD}{{\bf D}}
\newcommand{\PP}{{\bf P}}
\newcommand{\brl}{{<}l{>}}
\newcommand{\bari}{{\bf I}\!\!\!{\hspace{1pt}\scriptscriptstyle{{}^{-}}}}
\newcommand{\barj}{{\bf J}\!\!\!{\scriptscriptstyle{{}^{-}}}}
\newcommand{\barix}[1]{\bari^{{<}#1{>}}}
\newcommand{\barjx}[1]{\barj^{{<}#1{>}}}
\newcommand{\baril}[1]{{}^{#1}{\bari}{}^{\brl}}
\newcommand{\barjl}[1]{{}^{#1}{\barj}{}^{\brl}}
\newcommand{\lf}{\lfloor}
\newcommand{\rf}{\rfloor}
\newcommand{\la}{\lambda}
\newcommand{\boldla}{\mbox{\boldmath$\lambda$}}
\newcommand{\levi}{\varepsilon}
\begin{document}
\jl{6} 
\title{Gravitational waves from binary systems in circular orbits: 
Does the post-Newtonian expansion converge?}[Post-Newtonian
convergence of gravitational waves]
\author{Liliana E Simone\dag, Stephen W Leonard\ddag, Eric Poisson\ddag, 
and Clifford M Will\dag}
\address{\dag\ McDonnell Center for the Space Sciences\\
Department of Physics, Washington University\\
St. Louis, MO 63130, USA}
\address{\ddag\ Department of Physics, University of Guelph\\
Guelph, ON N1G 2W1, Canada} 

\begin{abstract}
Gravitational radiation can be expressed in terms of an infinite
series of radiative, symmetric trace-free (STF) multipole moments
which can be connected to the behavior of the source.  We consider a
truncated model for gravitational radiation from binary systems 
in which each STF mass and current moment of order $l$
is given by the lowest-order, Newtonian-like $l$-pole
moment of the orbiting masses; we neglect post-Newtonian
corrections to each STF moment.  Specializing to orbits which are
circular (apart from the radiation-induced inspiral), we find an
explicit infinite series for the energy flux
in powers of $v/c$, where $v$ is the orbital
velocity. 
We show that the series converges for all values $v/c < 2/{\rm e}$
when one mass is much smaller than the other, and $v/c < 4/{\rm e}$
for equal masses,
where $\rm e$ is the base of
natural logarithms.  These values include all physically relevant
values for compact binary inspiral.  
This convergence cannot indicate
whether or not the full series (obtained from the exact moments) will
converge.  But if the full series does not converge, our analysis
shows that this failure to converge does not originate from summing
over the Newtonian part of the multipole moments.
\end{abstract}
\maketitle

\section{Introduction}

The possibility of detection of gravitational waves from inspiralling
compact binaries using laser interferometric gravitational-wave
observatories such as the U.S. LIGO and the French-Italian VIRGO
projects has brought into sharp focus the accuracy of calculations of
gravitational waves using approximation methods.  The ability to measure
the source parameters using matched filtering of theoretical
templates against the tens of thousands of cycles observed in, say, a
double neutron-star inspiral, assumes that the templates are
sufficiently accurate, especially in the evolution of the phase of the waves, 
that the errors are smaller than the errors arising from noise in the
detectors \cite{finnchern,cutflan0}.
This may require knowledge of the damping of the orbit via
gravitational-radiation reaction (which determines the non-linear
evolution of the phase) that incorporates corrections to the
lowest-order quadrupole approximation as high as order $(v/c)^6$ 
\cite{cutflan,eric2}.
Corrections at $O[(v/c)^4]$ and $O[(v/c)^5]$ have recently been
calculated \cite{bdiww,bdi,ww,biww,luc}.

On the other hand, there is evidence that such post-Newtonian
expansions of gravitational radiation (weak-field, slow-motion
expansions in powers of $\epsilon^{1/2} \sim (v/c) \sim (Gm/rc^2)^{1/2}$)
do not converge rapidly, if at all.  For the case of a test body in
circular orbit around a black hole, perturbation calculations carried
to very high order in $v/c$ show slow convergence --- the coefficients
of successive powers of $v/c$ in the expansions grow alarmingly
quickly \cite{tag,tagsasak}.  
This could call into question the accuracy 
of any 
post-Newtonian approximation truncated at a finite order, and by
implication, the use of such approximations in templates used in data
analysis for LIGO and VIRGO.

To date, no method has been identified 
to study the convergence properties of the
post-Newtonian expansion rigorously and in generality, because of the
difficulty in generating higher-order corrections explicitly.  In the
case of post-Newtonian calculations for systems of arbitrary masses
\cite{bdi,ww}, 
the complexity of the
computations grows rapidly with each succeeding order.  Although the 
``third'' post-Newtonian order (3PN), corresponding to corrections at
$O[(v/c)^6]$ may be achievable, progress beyond that is unlikely.
Besides, 3PN order may be adequate from the data-analysis point of
view, given the expected noise characteristics of the advanced
LIGO/VIRGO detectors, provided one had some understanding of the convergence
properties, or could bound the errors neglected.

There is one situation in which the convergence properties can be
studied at all orders in $v/c$.  We call this the ``bare-multipole
truncation'',
and although it is unphysical, it may provide clues concerning
the convergence of the true (and physical) post-Newtonian expansion.  
This truncation is best discussed with reference to figure 1.  

Gravitational radiation can be expressed in terms of an infinite set
of radiative mass and current ``symmetric, trace-free (STF)'' multipole 
moments,
which can be related to multipole moments of the source
\cite{thorne80}.
For example, the rate of energy loss $\dot E$ and the gravitational waveform
far from the source $h^{ij}_{\rm TT}$ can be written
\numparts
\begin{eqnarray}
\fl \dot E = {G \over c^5} \sum_{l=2}^{\infty}
             \left( \alpha_l \,\,{^{(l+1)}\bari}^{a_1 \dots a_l}
		{^{(l+1)}\bari}^{a_1 \dots a_l} 
              +
         \beta_l \,\,{^{(l+1)}\barj}^{a_1 \dots a_l}
		{^{(l+1)}\barj}^{a_1 \dots a_l} \right) \,, 
\label{lumi1}\\
\fl h^{ij}_{\rm TT} = {G \over {Rc^4}} \sum_{l=2}^{\infty}
      \left( {4 \over {l!}} \,\,{^{(l)}\bari}^{ija_1 \dots a_{l-2}}
         \hat N^{a_1 \dots a_{l-2}} 
          + {{8l} \over {(l+1)!}} \levi_{pq(i} 
 	\,\,{^{(l)} \barj}^{j)pa_1 \dots a_{l-2}}
         \hat N^{qa_1 \dots a_{l-2}}  \right)_{\rm TT} \,,
\label{waveform}
\end{eqnarray}
\endnumparts
with
\begin{eqnarray}
\alpha_l = \frac{(l+1)\,(l+2)}{l(l-1)l!\,(2l+1)!!} \,,
\quad \beta_l = \frac{4l\,(l+2)}{(l-1)(l+1)!(2l+1)!!} \,, \label{coefbl}
\end{eqnarray}
where $\bari^{a_1 \dots a_l}$ and $\barj^{a_1 \dots a_l}$
are respectively the STF mass and current multipole moments 
of order $l$, $\hat N^{a_1 \dots a_l}$ denotes an $l$-dimensional 
product of unit vectors pointing from the source's center of
mass to the observer at a distance $R$, 
$\levi_{pqj}$ is the completely antisymmetric
Levi-Civita symbol, and the subscript ``TT'' denotes the 
transverse-traceless part. 
The indices on the left of the moments denote a number
of derivatives with respect to retarded time.
Summation over repeated indices is implied.

Each STF multipole moment is then expressed as a post-Newtonian (PN) expansion,
effectively in powers of $\epsilon^{1/2}$, starting 
with a lowest-order multipole
moment given in general by
\numparts
\begin{eqnarray}
\barix{l}_{(0)} &=& \left( \sum_A m_A x_A^{i_1}x_A^{i_2}\dots x_A^{i_l}
	\right)_{\rm STF} \,, \\
\barjx{l}_{(0)} &=& \left( \levi_{i_1 ab} \sum_A m_A 
    x_A^a v_A^b x_A^{i_2}\dots x_A^{i_l}
	\right)_{\rm STF} \,, 
\end{eqnarray}
\endnumparts
where $m_A$, $x_A^i$, and $v_A^i$ are the suitably defined mass, position,
and velocity of each body, and 
the superscript notation $<l>$ is short-hand for the $l$ indices.

Figure 1 represents these STF moments in an array with multipole order
$l$ increasing horizontally beginning with $l=2$ (for each $l$,
$\barix{l}$ and $\barjx{l-1}$ are grouped together), and with PN order for
each multipole moment
increasing downward.  (Note that $\barix{0}$, $\barix{1}$, and
$\barjx{1}$ are related to the non-radiative total mass, center of mass, and
total angular momentum of the system.)  It is straightforward to show
that the first-order PN correction to each multipole moment
is $O(\epsilon)$,
followed by a ``tail'' correction at $O(\epsilon^{3/2})$, 
then $O(\epsilon^2)$, and so on.  We ignore finite-size effects, such as the
effects of
spin, in this discussion.  There is evidence from black-hole
perturbation studies that the PN expansions become non-analytic at
high enough order, with the appearance of $\ln \epsilon$ terms
\cite{tag,tagsasak}.

Because each multipole moment
is differentiated with respect to retarded time
a number of times corresponding to its multipole order, and because
$d/dt \sim \epsilon^{1/2}/r_A$, a specific moment
contributes to $h^{ij}_{\rm TT}$ and to $\dot E$ at a PN order related
to its multipole order. For example, a determination of $\dot
E$ to 2PN order requires knowledge only of the first three columns of 
figure 1,
($l=2, \; 3,\; 4$), together with corrections through 2PN order of
the quadrupole STF mass moment ($\barix{2}$), 
the 1PN corrections of the octopole mass
and quadrupole current moments ($\barix{3}$, $\barjx{2}$), 
and the lowest-order hexadecapole 
mass moment ($\barix{4}$) and octopole current moment ($\barjx{3}$).  
These correspond to the shaded
region in figure 1.  For the waveform at 2PN order, on the other hand,
the moments and their corrections indicated by the dark border are
required.  The difference derives from the fact that the waveform is linear 
in the moments, while the energy flux is quadratic.

The ``bare-multipole 
truncation'' 
consists in keeping only the first
row of figure 1, 
but keeping {\it all} orders in $l$.  This is not a consistent
PN approximation to either the waveform or the energy loss. 
Moreover, this truncation does not correspond to any sort of physical
model for the system, such as one in which 
non-gravitational forces are responsible for keeping the two bodies in
orbital motion.  This can be seen as follows:  Consider the 
$O(\epsilon)$ corrections to a particular multipole moment.  These
contain terms of the form $Gm/rc^2$, which can be thought of as the
gravitational contribution to the effective stress-energy tensor, and
terms of the form $(v/c)^2$, which can be thought of as special
relativistic corrections in the matter contribution to the effective
stress-energy tensor.  Now, it is clear that the first set of
correction terms would be discarded in a non-gravitational model for
the source motion, although it would presumably be replaced by
something else (related the stresses responsible for maintaining the
orbit).  However, it is also clear that the second set of terms cannot
be discarded in any sort of physical model for the source.  The fact
that both sets are discarded in our truncation clearly makes such a
procedure unphysical.

Although it is unphysical, the bare-multipole truncation may
illustrate some of the convergence properties of the full formulae.
For two-body systems, we show that the bare moments have the general
form
\begin{eqnarray}
\barix{l}_{(0)}=\mu r^l\left. f_l(\eta)\, \SS^l_0 \right|_{\rm TF}\,,
\nonumber\\
\barjx{l}_{(0)}=\mu r^l\,v\,f_{l+1}(\eta)
            \left[{\bf \hat L}\circ \SS^{l-1}_0 \right]_{\rm TF}\,,
\label{bare}
\end{eqnarray}
where $\mu = m_1m_2/(m_1+m_2)$ is the reduced mass, $\eta=\mu/(m_1+m_2)$,
$\SS^l_0$ is a symmetrized product of $l$ unit radial vectors directed from 
body 2 to body 1, $\hat{\bf L}$ 
is a unit vector in the direction of the orbital angular momentum, $r$
and $v$ are the 
magnitudes of the relative separation and orbital velocity,
respectively, and 
$\circ$ denotes a symmetrized product.  The function $f_l(\eta)$ is given
by 
\begin{eqnarray}
f_l(\eta)=\left[\rho^{l-1}+(-1)^l\right]/
          {(1+\rho)^{l-1}} \,,\\
\rho \equiv m_2/m_1 = \frac{1}{2\eta}\left[1-2\eta - \sqrt{1-4\eta}\right] \,,
\label{fsubl}
\end{eqnarray}
where we choose the convention $m_2 \le m_1$, so that 
$0 < \rho \le 1$.

Specializing to quasi-circular orbits with angular velocity $\Omega
\equiv v/r$,
we find, after considerable
manipulations of STF tensors (Section 3), the closed-form result
\begin{eqnarray}
\dot E = \dot E_Q \left \{ 1+ \sum_{l=3}^\infty B_l f_l^2 (\eta)
\left({v \over c}\right)^{2l-4} \right \} \;,
\label{edotfinal}
\end{eqnarray}
where $\dot E_Q$ represents the quadrupole approximation, 
\begin{eqnarray}
\dot E_Q = {{32G} \over {5c^5}} \mu^2 v^4 \Omega^2 \,. 
\end{eqnarray}
The coefficients $B_l$ are given explicitly by \eref{bsubl}.
It is interesting to note here that \eref{edotfinal} does not depend
on any assumption of an equation of motion for the bodies; only the
existence of a quasi-circular orbit with angular velocity $\Omega$ is
assumed.

A completely equivalent series, 
applicable in the test mass limit ($\eta \to
0$, $|f_l(\eta)| \to 1$) 
may be derived using black-hole perturbation theory (see Section 4 for
details).
The coefficients $B_l$ resulting from that
method have a different, albeit numerically equivalent representation,
given by (\ref{e.6}, \ref{e.7}).  
Most importantly, they can easily be shown to 
have the form, for large $l$,
\begin{eqnarray}
B_l \sim {5 \over 64} \left ( {l \over {\pi}} \right )^{1/2} 
 \left ( {{\rm e} \over 2} \right )^{2l} 
\Bigl[1 + O\bigl(l^{-1}\bigr)\Bigr]  \,,
\end{eqnarray}
where ${\rm e}$ is the base of natural logarithms.  

By applying the standard Cauchy ratio test, we find that
the radius of convergence of the sequence is given by
\begin{eqnarray}
\left({v \over c}\right)_{\rm converge} 
   = {4 \over { {\rm e}\,(1+ \sqrt{1-4\eta} ) }} \;. 
\end{eqnarray}
The series thus
converges for values of $v/c$ less than a critical value ranging from
0.74 ($\eta=0$) to 1.47 ($\eta=1/4$).  These values
encompass all physically relevant values in binary inspiral, until
the final coalescence phase.

The infinite series \eref{edotfinal} can be used to study the error made in 
truncating the method at a given PN order. (By this we mean a truncation
of the sum appearing in Eq.~\eref{edotfinal} at the value of $l$ 
corresponding to the specified PN order. For example, a truncation at
2PN order involves keeping only the $l=3$ and $l=4$ terms.)  Figure 2
shows, as a function 
of $v$, the fractional difference between a given PN truncation and  the 
full series.  Notice that, at $v \approx 0.4$, corresponding to the 
innermost stable orbit for a test-body orbiting a black hole, the errors at 
2PN and 3PN order are 2.1 and 0.6 percent, respectively.  For equal
masses,
odd-numbered mass moments and even-number current moments vanish, thus
there are no terms in $\dot E$ at odd-PN order in the bare-multipole
truncation.
This is partly responsible for the improved
convergence in the equal-mass case: at $v \approx 0.4$, the fractional
difference between the 2PN/3PN approximation and the full series is $2
\times 10^{-5}$.  

The remainder of this paper provides details of the calculations.  In 
Section 2, we describe the bare-multipole 
truncation
and set up our 
definitions and conventions.  Section 3 derives the energy flux, and 
discusses the convergence of the series.  In Section 4,
we describe the analogous approach using black hole perturbation theory, 
and find an analytic expression for the radius of convergence.
Section 5 gives concluding remarks.  A number of technical details are 
relegated to Appendices.  

\section{Bare-multipole truncation for gravitational radiation
\label{mysect}}

We describe the motion of a system of two bodies whose size is negligible
compared to their relative separation. We work in a 
coordinate system whose origin is the center of 
mass, and define the relative position vector 
${\bf x}\equiv{\bf x}_1-{\bf x}_2$,
where ${\bf x}_1$ and ${\bf x}_2$ are the positions of each body. 
If the masses of the bodies are $m_1$ and $m_2$, we define 
$m\equiv m_1+m_2$, $\mu\equiv m_1\,m_2/m$, $\eta\equiv\mu/m$, and 
$\delta m\equiv m_1-m_2$. Also, 
$r\equiv |{\bf x}|$ and ${\bf v}\equiv\dot{\bf x}$.  Henceforth,
we use units in which $G=c=1$.

The radiative multipole moments
in \eref{lumi1} and \eref{waveform} 
are expanded in powers 
of $\epsilon^{1/2}
\sim v \sim (m/r)^{1/2} \sim \dot r$, in the generic form
\begin{eqnarray}
\barix{l} \sim \barix{l}_{(0)} (1 + \epsilon + \epsilon^{3/2} + \epsilon^2 +
\dots ) \,,
\end{eqnarray}
with a similar expansion for $\barjx{l}$, where the $\epsilon^{3/2}$
contribution signals the first appearance of the effects of 
radiative ``tails'', caused by backscatter of the waves off the
background spacetime curvature.  In calculating $\dot
E$, one takes each multipole moment of order $l$, differentiates it $l+1$
times with respect to retarded time, and squares the result
(contracting on all indices).
Because the $l$-pole moment is 
differentiated once more than the $(l-1)$-pole moment, its leading
contribution to $\dot
E$ is of order $v^2$ higher than that of the $(l-1)$-pole moment.  
Additionally, 
the current multipole moments
are proportional to the angular momentum of the system,
through a factor $\levi_{iab}\, x^a\,v^b$, which adds an extra velocity 
$v^b$ to their expressions; consequently the contribution of a current 
multipole moment
is $O(v^2)$ higher than that of the mass multipole moment of the same rank.
As a result, \eref{lumi1} is a post-Newtonian expansion 
of the luminosity in powers of $\epsilon$.   Similar 
considerations apply to the
expressions for the gravitational waveform. 

Explicit expressions for the 
radiative mass and current 
multipole moments for general two-body systems are now known 
sufficiently accurately to calculate the energy flux to 5/2PN order beyond
the quadrupole approximation \cite{luc}, and the waveform to 2PN order
\cite{biww}.  By way
of illustration, we quote here the moments sufficient to calculate the
2PN energy flux \cite{bdi,ww}:
\begin{eqnarray}
\fl{\bari}^{ij} = 
\mu\Biggl\{\left[1+\frac{29}{42}\,(1-3\eta)\,v^2
 -\frac{1}{7}\,(5-8\eta)\left(\frac{m}{r}\right)\right] x^i\,x^j
           \nonumber\\
 -\frac{4}{7}\,(1-3\eta)\,r\,\dot r\,x^i\,v^j
 +\frac{11}{21}\,(1-3\eta)\,r^2\,v^i\,v^j \nonumber\\
\hspace{-0.7in}        
        +x^i\,x^j
\biggl[\frac{1}{504}\left(253-1835\,\eta +3545\,\eta^2\right) v^4
+\frac{1}{756}\left(2021-5947\,\eta-4883\,\eta^2\right) v^2
\left(\frac{m}{r}\right) \nonumber\\
\hspace{-0.7in}        
        -\frac{1}{252}\left(355+1906\,\eta-337\,\eta^2\right) 
\left(\frac{m}{r}\right)^2
-\frac{1}{756}\left(131-907\,\eta+1273\,\eta^2\right) 
{\dot r}^2\left(\frac{m}{r}\right)\biggr] \nonumber\\
\hspace{-0.7in}        
        + r^2\,v^i\,v^j\biggl[\frac{1}{189}
   \left(742-335\,\eta-985\,\eta^2\right)
   \left(\frac{m}{r}\right)
   +\frac{1}{126}\left(41-337\,\eta+733\,\eta^2\right) v^2 
 \nonumber\\
   +\frac{5}{63}\left(1-5\eta+5\eta^2\right){\dot r}^2\biggr]
 \nonumber\\
\hspace{-0.7in}        
        - r\dot r\, v^i x^j
   \biggl[\frac{1}{378}\left(1085-4057\,\eta-1463\,\eta^2\right)
     \left(\frac{m}{r}\right)
   +\frac{1}{63}\left(26-202\,\eta+418\,\eta^2\right) v^2\biggr]
 \Biggr\}_{{\rm STF}} \nonumber\\
\hspace{-0.7in}        
    + {\bari}^{ij}_{\rm Tail} \,,
\nonumber\\
\fl {\bari}^{ijk} = -\mu\frac{\delta m}{m}
 \Biggl\{ \left[1+\frac{1}{6}\,(5-19\,\eta)\,v^2
 -\frac{1}{6}\,(5-13\,\eta)
  \left(\frac{m}{r}\right)\right]\,x^i\,x^j\,x^k  
       \nonumber\\
 +(1-2\eta)(r^2\,v^i\,v^j\,x^k-r\,\dot r\,v^i\,x^j\,x^k) 
  \Biggr\}_{{\rm STF}} \,,
\nonumber\\ 
\fl {\bari}^{ijkl}=\mu\left(1-3\eta\right) (x^i\,x^j\,x^k\,x^l)_{{\rm STF}}\,,
\nonumber\\
\fl {\barj}^{ij} = -\mu\frac{\delta m}{m} 
 \Biggl\lgroup \levi_{iab}\, 
 \biggl\{ \left[1+\frac{1}{2}\,(1-5\eta)\,v^2
 +2\,(1+\eta)\left(\frac{m}{r}\right)\right]x^jx^av^b \nonumber \\
         +\frac{1}{28}\frac{d}{dt}\left[(1-2\eta) 
  (3r^2v^j-r\dot rx^j)\,x^av^b\right] 
\biggr\} \Biggr\rgroup_{{\rm STF}} \,,
\nonumber\\
\fl {\barj}^{ijk} = \mu\left(1-3\eta\right)
(\levi_{iab}\, x^a\,v^b\,x^j\,x^k)_{{\rm STF}}\,. 
\label{2pnmults}
\end{eqnarray}
Explicit expressions for the ``Tail'' term may be found in \cite{bdi,ww},
for example.

We now specialize to quasi-circular orbits, {\it i.e.} orbits 
which are circular,
apart from the slow inspiral caused by radiation damping
(at sufficiently high PN order the non-circularity of the
orbits must be taken into account).  We thus
approximate: $\dot r=0$, ${\bf x}=r{\bf \hat n}$, 
${\bf v}=v\hat{\boldla}$, where ${\bf \hat n}$ and $\hat
{\boldla}$ are unit vectors in 
the radial and tangential directions, respectively,
and ${\bf\hat n}\times\hat{\boldla}={\bf\hat L}$.
We define the following symmetric tensors~(see~\ref{appa}): 
\begin{equation}
\SS_k^{l-k}\equiv\frac{1}{\cal N}\sum_{\Pi_{\cal N}}
	  \left[\hat\la^{i_1}\,\hat\la^{i_2}\dots\hat\la^{i_k}\,
                \hat n^{i_{k+1}}\,\hat n^{i_{k+2}}\dots \hat n^{i_l} \right],
\label{sym-tensors}
\end{equation}
where the sum is over all possible distinct permutations ${\cal N} = 
{l \choose k}$ of the set of indices
$\{i_j\}$.  The STF tensors 
$\barix{l}$ and $\barjx{l}$ can then be expressed in terms of
these symmetric tensors, and the post-Newtonian terms can be
converted, via the appropriate equations of motion, into terms
involving $v$ alone.  The result is, schematically,
\begin{eqnarray}
\barix{2}=\mu r^2\bigl[
     (1+\alpha_1 v^2 +\alpha_2 v^4)\SS^2_0  +v^2(1+\alpha_3 v^2) \SS^0_2
      \bigr]_{\rm STF} +\barix{2}_{\rm Tail} \,, \nonumber\\
\barix{3}=\mu r^3\left(-\frac{\delta m}{m}\right)
      \left[(1+\beta_1 v^2)\SS^3_0+\beta_2v^2\SS^1_2
      \right]_{\rm STF} \,, \nonumber\\
\barix{4}=\mu r^4 \left[ \gamma_1\SS^4_0\right]_{\rm STF}\,,\nonumber\\
\barjx{2}=\mu r^2v \left( -\frac{\delta m}{m} \right) (1+\delta_1v^2)
      \left[{\bf\hat L}\circ\SS^1_0 \right]_{\rm STF} \,, \nonumber\\
\barjx{3}=\mu r^3v\epsilon_1
      \left[{\bf\hat L}\circ \SS^2_0
      \right]_{\rm STF}\,,
\end{eqnarray}
where the coefficients $\alpha_i$, $\beta_i$, etc., result from
combining expressions in \eref{2pnmults}, 
and depend only on $\eta$; $\circ$ denotes a symmetrized product, 
defined by
\begin{equation}
{\bf\hat L}\circ \SS^{T-1}_0
\equiv \hat L^{(i_1}\hat n^{i_2}\cdots \hat n^{i_T )}
\equiv \frac{1}{T}\sum_{\Pi_{T}} \hat L^{i_1}\hat n^{i_2}\cdots \hat
n^{i_T},
\label{symmet}
\end{equation}
where $\Pi_{T}$ are all the $T$ distinct 
permutations of the indices $\{i_j\}$.
We now restrict these expressions to the leading order in $v$, and
denote them the ``bare'' multipole moments $\barix{l}_{(0)}$ and
$\barjx{l}_{(0)}$, as follows 
\begin{eqnarray}
\barix{2}_{(0)}=\mu r^2 \left. \SS^2_0 \right|_{\rm STF}\equiv 
        \mu r^2 \left. f_2(\eta) \SS^2_0 \right|_{\rm STF},\nonumber
\\
\barix{3}_{(0)}=\mu r^3 \left(-\frac{\delta m}{m}\right)
        \left. \SS^3_0 \right|_{\rm STF}\equiv 
        \mu r^3 \left. f_3(\eta) \SS^3_0 \right|_{\rm STF},\nonumber
\\
\barix{4}_{(0)}=\mu r^4 (1-3\eta)
        \left. \SS^4_0 \right|_{\rm STF}\equiv 
        \mu r^4 \left. f_4(\eta) \SS^4_0 \right|_{\rm STF},\nonumber
\\
\vspace{2mm}
\barjx{2}_{(0)}=\mu r^2v\left(-\frac{\delta m}{m}\right)
      \left[{\bf\hat L}\circ \SS^1_0 \right]_{\rm STF}\equiv
      \mu r^2vf_3(\eta)
      \left[{\bf\hat L}\circ \SS^1_0 \right]_{\rm STF}, \nonumber
\\
\barjx{3}_{(0)}=\mu r^3v(1-3\eta)
      \left[{\bf\hat L}\circ \SS^2_0 \right]_{\rm STF}\equiv
      \mu r^3vf_4(\eta)
      \left[{\bf\hat L}\circ \SS^2_0 \right]_{\rm STF},
\label{IJ234}
\end{eqnarray}
where the function $f_l(\eta)$ is given by \eref{fsubl}
and is plotted in figure 3 for different
values of $\eta$ within the possible range $(0,0.25]$. Note that
$\left|f_l(\eta)\right|<1$, for all $\eta\ne0$; $|f_l(0)|=1$, for all
$l$ and $f_l(0.25)=0$ for odd $l$ (see~\ref{appb}).
Equations \eref{IJ234} suggest, and Appendix B confirms, that 
the general forms for $\barix{l}_{(0)}$ and $\barjx{l}_{(0)}$ 
at leading order are 
given by \eref{bare}.
Notice that the original expressions for $\barix{l}_{(0)}$ 
and $\barjx{l}_{(0)}$
were expressed in harmonic coordinates. However, these expressions
are
also valid in Schwarzschild coordinates, as the transformation between
coordinate systems, both for $r$ and $v$, introduces terms that 
contribute only higher-order corrections.

Finally, we explicitly extract the traces in the 
previous expressions, and get $\barix{l}_{(0)}$ and $\barjx{l}_{(0)}$
in terms of the symmetric tensors $\SS_k^{l-k}$ 
only,
\numparts
\begin{eqnarray}
\fl\barix{l}_{(0)}= \mu r^lf_l(\eta)\sum_{j=0}^{\left\lf\frac{l}{2}\right\rf}
   \frac{(-1)^j{l\choose j}{l\choose{2j}}}{{{2l}\choose{2j}}}
   \left(\DD^{(j)}\circ \SS^{l-2j}_0\right) \,,   
\label{stfsa}  \\
\vspace{3mm}
\fl\barjx{l}_{(0)}= \mu r^l vf_{l+1}(\eta)\sum_{j=0}^{\left\lf\frac{l-1}{2}
   \right\rf}
   \frac{(-1)^j{l\choose j}{l\choose{2j}}}{{{2l}\choose{2j}}}
   \frac{(l-2j)}{l}
   \left(\DD^{(j)}\circ\left({\bf\hat L}\circ \SS^{l-2j-1}_0\right)\right)
\,.
\label{stfsb}
\end{eqnarray}
\endnumparts
Here $\lf a/b\rf$ denotes the integer part of $a/b$, and 
$\DD^{(j)}$ is a symmetrized product of $j$ Kronecker deltas, given
by
\begin{eqnarray}
\DD^{(j)}=\delta^{(i_1 i_2}\delta^{i_3 i_4} \cdots \delta^{i_{2j-1},i_{2j})} 
   =\frac{1}{{\cal N}(j,2)}
          \sum_{\Pi_{{\cal N}(j,2)}}\delta^{i_1 i_2}\delta^{i_3 i_4}
          \cdots \delta^{i_{2j-1},i_{2j}} \,,
\label{delta-op}
\end{eqnarray}
where
\begin{eqnarray}
{\cal N}(j,2) \equiv \frac{1}{j!}{2j\choose 2}
                {{2j-2}\choose 2}\cdots{2\choose 2}
         = (2j-1)!!
\label{delta-num}
\end{eqnarray}  
is the number of distinct products of deltas.
The factor $(l-2j)/l$ in~\eref{stfsb} comes from taking into
account the fact that
the unit vector ${\bf\hat L}$ is perpendicular 
to ${\bf\hat n}$ when applying traces to the
tensor ${\bf\hat L}\circ \SS^{l-2j-1}_0$.

Notice that each term in \eref{stfsa} and \eref{stfsb} has rank 
$l-2j$, varying 
for each $j$ in the sum. This fact complicates considerably the process of
multiplying two of these objects, as prescribed in \eref{lumi1}. In
the next section we will
describe how to simplify this calculation.

\section{Energy flux in the bare-multipole 
truncation
}

We now express the
energy flux approximately in terms of the
bare multipole moments 
\begin{eqnarray}
\dot E \approx  \sum_{l=2}^{\infty}
             \left( \alpha_l {^{(l+1)}}\barix{l}_{(0)} \cdot
                {^{(l+1)}}\barix{l}_{(0)}
              +
         \beta_l {^{(l+1)}}\barjx{l}_{(0)} \cdot
                {^{(l+1)}}\barjx{l}_{(0)} \right) \,,
\label{lumi2}
\end{eqnarray}
where
the notation $(\cdot)$ stands for the inner product that saturates all 
the indices $\{i_l\}\equiv\{i_1,i_2,\dots,i_l\}$ of the two tensors.
As we are dealing henceforth with bare multipole moments, we drop the
subscript $(0)$ notation.

In order to simplify the 
product of derivatives in each of the sums of
\eref{lumi2}, we first write:
\begin{eqnarray}
\fl\baril{(l+1)}\cdot\baril{(l+1)} =
\frac{d}{dt}\left[\baril{(l)}\cdot\baril{(l+1)}\right]
      -\baril{(l)}\cdot\baril{(l+2)}
		     \nonumber\\
\qquad =\frac{d}{dt}\left[\qquad\right]
   -\Biggl\{\frac{d}{dt}\left[\baril{(l-1)}\cdot\baril{(l+2)}\right]
   -\baril{(l-1)}\cdot\baril{(l+3)}\Biggr\}
		     \nonumber\\
\qquad =\frac{d}{dt}\left[\qquad\right] +\baril{(l-1)}\cdot\baril{(l+3)}
		     \nonumber\\
\qquad   \qquad\qquad\dots    \nonumber\\
\qquad =\frac{d}{dt}\left[\qquad\right] 
           +(-1)^{l+1}\baril{}\cdot\baril{(2l+2)}.
\end{eqnarray}

The first term is the total derivative of several terms, 
each of which is the product of an even and an odd 
number of derivatives of 
$\baril{}$, respectively.  Since each time derivative converts 
${\bf\hat n}$ into $\hat{\boldla}$, each term is thus 
a product of a tensor with an even number of $\hat{\boldla}$ 
unit vectors by a tensor with an odd number of $\hat{\boldla}$'s;
consequently, every term in the bracket is null, because ${\bf \hat n}
\cdot \hat{\boldla} =0$. The same can be done
with the magnetic terms, so that we may write
\begin{eqnarray}
\baril{(l+1)}\cdot\baril{(l+1)}&= (-1)^{l+1}\baril{}\cdot\baril{(2l+2)},
\nonumber\\
\barjl{(l+1)}\cdot\barjl{(l+1)}&= (-1)^{l+1}\barjl{}\cdot\barjl{(2l+2)}.
\end{eqnarray}

Now, $\baril{}$ and $\barjl{}$ are trace-free tensors, as are their
derivatives.  But the terms in $\baril{}$ that 
contain Kronecker deltas, such as 
$\delta^{i_1 i_2}({\rm Rest})^{i_3\cdots i_l}$, are just trace operators 
on the subspace $\{i_1,i_2\}$; so their product with the time-differentiated
STF tensor 
$\baril{(2l+2)}$ will vanish. Thus, the only 
non-vanishing term is the symmetric product of $l$ unit vectors 
$\left[\hat n^{i_1}\,\hat n^{i_2}\cdots \hat n^{i_l}\right]$
multiplied by $\baril{(2l+2)}$.  A similar argument applies to the
current moments.  The result is
\begin{eqnarray}
\baril{(l+1)}\cdot\baril{(l+1)}
   &= (-1)^{l+1}\mu r^l\left[\hat n^{i_1} \hat n^{i_2}\cdots \hat n^{i_l}\right]
   \cdot\baril{(2l+2)} \nonumber\\
   &= (-1)^{l+1} \mu r^l \SS^l_0 \cdot\baril{(2l+2)} \,,
\nonumber\\
\barjl{(l+1)}\cdot\barjl{(l+1)}
   &= (-1)^{l+1} \mu r^l v
   \left[\hat L^{(i_1} \hat n^{i_2} \hat n^{i_3}\cdots \hat n^{i_l)}\right]
   \cdot\barjl{(2l+2)} \nonumber\\
   &= (-1)^{l+1} \mu r^l v\left[{\bf\hat L}\circ \SS^{l-1}_0\right]
       \cdot\barjl{(2l+2)}.
\label{prod}
\end{eqnarray}

We now need an expression for the $(2l+2)$-\-time-\-derivative of $\baril{}$
and $\barjl{}$.  These tensors
(\ref{stfsa}, \ref{stfsb}) are 
sums of products of constants with respect to time (like $\DD^{(j)}$ 
and ${\bf\hat L}$) with
a symmetric tensor $\SS^{l-2j}_0$, which has only radial components.
We will concentrate first on calculating the $2p\equiv 2l+2$ 
derivatives of a symmetric tensor $\SS^T_0$ where 
$T=l-2j=l,l-2,l-4,\cdots$. 

The derivative of any symmetric product of unit vectors, radial and/or 
tangential is (see~\ref{appa1})
\begin{equation}
{d \over {dt}} \left(\SS^{l-k}_k\right)=
\Omega\left[(l-k) \SS^{l-k-1}_{k+1}-k \SS^{l-k+1}_{k-1}\right].
\label{oneder}
\end{equation}
Thus, even if we have only radial components in $\SS^T_0$, tangential 
components will appear after differentiation. Notice that~\eref{oneder} is
the same rule as the derivative of a product of 
$k$ cosine and $l-k$ sine functions
\begin{eqnarray}
\frac{d}{dx}\left(\sin^{l-k}\Omega x\,\cos^k\Omega x \right) &=
\Omega\Biggl[(l-k)\left(\sin^{l-k-1}\Omega x\,\cos^{k+1}\Omega x\right)
              \nonumber\\
&\qquad -k\left(\sin^{l-k+1}\Omega x\,\cos^{k-1}\Omega x\right) \Biggr]\,.
\label{der-sinecosine}
\end{eqnarray}
We may thus simplify the calculation considerably by associating
each symmetric tensor with a product of
sines and cosines in the form
\begin{equation}
\SS^{l-k}_k\quad\longleftrightarrow\quad
\sin^{l-k}\Omega x\,\cos^k\Omega x.
\end{equation}

Any higher-order derivative of $\SS^{l-k}_k$, expanded in terms of others
$\SS^{l-k'}_{k'}$, will have the same coefficients as the corresponding
derivative of the
analogous product of sines and cosines, expanded in products of sines and 
cosines, provided that we {\em can} express
such a derivative in terms of the equivalent products of sines and cosines
(in~\ref{appc} we show that there is a subtlety in this procedure).

We are interested in the ($2p$)-\-derivative of $\SS^T_0$, which we 
associate with $\sin^T \Omega x$, with $T=l,l-2,l-4,\cdots$. 
To get a closed 
expression, we first expand $\sin^T \Omega x$ as a sum of sines of multiple
angles \cite{grad}:
\begin{eqnarray}
\fl\sin^T \Omega x = \frac{1}{2^{T-1}}\left\{
  \sum^{\left\lf\frac{T-1}{2}\right\rf}_{k=0}
     (-1)^{\left\lf\frac{T}{2}\right\rf-k}\,{T\choose k}
	  \cos\Omega(T-2k)x 
	  +\frac{1}{2}\,{T\choose{\left\lf\frac{T}{2}\right\rf}}\right\}\,,
   	  \quad \mbox{if }T=2n,\nonumber\\
\fl\sin^T \Omega x = \frac{1}{2^{T-1}}
  \sum^{\left\lf\frac{T-1}{2}\right\rf}_{k=0}
     (-1)^{\left\lf\frac{T-1}{2}\right\rf+k}\,{T\choose k}
	  \sin\Omega(T-2k)x \,,
   	  \quad\qquad \mbox{if }T=2n-1. 
\label{sines}
\end{eqnarray}

We then take the $2p$ derivative of~\eref{sines}:
\begin{eqnarray}
\fl ^{(2p)}(\sin^T \Omega x) = \frac{(-1)^p\Omega^{2p}}{2^{T-1}}
  \sum^{\left\lf\frac{T-1}{2}\right\rf}_{k=0}
     (-1)^{\left\lf\frac{T}{2}\right\rf-k}\,{T\choose k}
	  \,(T-2k)^{2p}\,\cos\Omega(T-2k)x \,,\, \mbox{if }T=2n,
\nonumber\\
\fl ^{(2p)}(\sin^T \Omega x) = \frac{(-1)^p\Omega^{2p}}{2^{T-1}}
  \sum^{\left\lf\frac{T-1}{2}\right\rf}_{k=0}
     (-1)^{\left\lf\frac{T-1}{2}\right\rf+k}\,{T\choose k}
	  \,(T-2k)^{2p}\,\sin\Omega(T-2k)x \,,\nonumber\\
   	  \qquad\qquad\qquad\qquad\qquad 
   	  \qquad\qquad\qquad\qquad\qquad \mbox{if }T=2n-1.
\label{der-sines}
\end{eqnarray}
Now we can express $\cos\Omega(T-2k)x$ and $\sin\Omega(T-2k)x$ 
back in terms of powers of sines and cosines, to recover terms analogous to the
symmetric products $\SS_k^{l-k}$:
\begin{eqnarray}
\fl \cos\Omega(T-2k)x  =  \sum^{\left\lf\frac{T}{2}\right\rf-k}_{q=0}
     (-1)^q\,{{T-2k}\choose{2q}}\,\sin^{2q}\Omega x\,
	      \cos^{(T-2k-2q)}\Omega x  \,, \quad \mbox{if }T=2n,
\nonumber\\
\fl \sin\Omega(T-2k)x  =  \sin\Omega x
     \sum^{\left\lf\frac{T-1}{2}\right\rf-k}_{q=0}
     (-1)^q\,{{T-2k}\choose{2q+1}}\,\sin^{2q}\Omega x\,
	      \cos^{(T-2k-2q-1)}\Omega x  \,,\nonumber\\
   	      \qquad\qquad\qquad\qquad\qquad 
   	      \qquad\qquad\qquad\qquad\qquad \mbox{if }T=2n-1.
\label{multiples}
\end{eqnarray}
Combining \eref{der-sines}
and \eref{multiples}, one can get the $2p$-\-derivative 
of a product of $T$ sines. The $2p$-\-derivative of $\SS_0^T$ will
then have a similar expansion, with identical coefficients:
\numparts
\begin{eqnarray}
\fl {}^{(2p)}(\SS_0^T) = \frac{(-1)^p\Omega^{2p}}{2^{T-1}}
  \sum^{\left\lf\frac{T-1}{2}\right\rf}_{k=0}
     (-1)^{\left\lf\frac{T}{2}\right\rf-k}{T\choose k}
	  (T-2k)^{2p}      \nonumber\\
  \quad\qquad\times\sum^{\left\lf\frac{T}{2}\right\rf-k}_{q=0}
     (-1)^q{{T-2k}\choose{2q}}\SS_{T-2k-2q}^{2q} \,,\qquad
   	  \quad \mbox{if }T=2n,
\label{der-symmeta}\\
\vspace{3mm}
\fl {}^{(2p)}(\SS_0^T) = \frac{(-1)^p\Omega^{2p}}{2^{T-1}}
  \sum^{\left\lf\frac{T-1}{2}\right\rf}_{k=0}
     (-1)^{\left\lf\frac{T-1}{2}\right\rf+k}{T\choose k}
	  (T-2k)^{2p}        \nonumber\\
  \quad\qquad\times\sum^{\left\lf\frac{T-1}{2}\right\rf-k}_{q=0}
     (-1)^q{{T-2k}\choose{2q+1}}\SS_{T-2k-2q-1}^{2q+1} \,,
   	  \quad \mbox{ if }T=2n-1.
\label{der-symmetb}
\end{eqnarray}
\endnumparts

However, this expansion of ${}^{(2p)}(\SS_0^T)$ is not yet what we need. 
Notice that the derivative rule~(\ref{oneder}) preserves the rank 
of the tensors involved:
${\rm rank}(\SS^{l-k}_k)={\rm rank}(\SS^{l-k-1}_{k+1})
={\rm rank}(\SS^{l-k+1}_{k-1})$. This is no longer true 
for (\ref{der-symmeta}, \ref{der-symmetb}),
where each term is of rank $T-2k$, different for each $k$. This is an 
artifact created by expanding the $\sin^{T}\Omega x$ in terms of sines and 
cosines of multiple angles, which we resolve in~\ref{appc}. 
The product of these expressions with $\SS_0^l$ or with 
${\bf\hat L}\circ\SS_0^{l-1}$ in \eref{prod}, selects only the
terms of the form $\SS_0^l$ in equations (\ref{der-symmeta},
\ref{der-symmetb}).
The expressions simplify considerably (see~\ref{appc}), giving
the same result for even or odd $T$. Using~(\ref{c-result}), we get 
the following product, valid for any $T,T'$ (even or odd) and for any
$p$:
\begin{eqnarray}
\SS_0^{T'}\cdot {}^{(2p)}\SS_0^T = {\cal A}(T,p)
\left(\SS_0^{T'}\cdot \SS_0^T \right) \,,
\end{eqnarray}
where
\begin{eqnarray}
{\cal A}(T,p) \equiv \frac{(-1)^p\Omega^{2p}}{2^{T-1}}
  \sum^{\left\lf\frac{T-1}{2}\right\rf}_{k=0} {T\choose k} (T-2k)^{2p}
\,.
\end{eqnarray}
As the angular momentum ${\bf\hat L}$ or the Kronecker delta operators
$\DD^{(j)}$ are constants with respect to time derivatives, we also find 
that
\begin{eqnarray}
\fl \SS_0^{T'}\cdot \left[\DD^{(j)}\circ {}^{(2p)}\SS_0^T\right]
     = {\cal A}(T,p) 
        \left(\SS_0^{T'}\cdot \left[\DD^{(j)}\circ \SS_0^T\right] \right)
\,,
\nonumber\\
\fl \left[{\bf\hat L}\circ\SS_0^{T'}\right] \cdot
   \left[\DD^{(j)}\circ\left[{\bf\hat L}\circ {}^{(2p)}\SS_0^T\right]\right]
  = {\cal A}(T,p) 
     \left(\left[{\bf\hat L}\circ\SS_0^{T'}\right] \cdot 
     \left[\DD^{(j)}\circ\left[{\bf\hat L}\circ\SS_0^T\right]\right] \right)
\,.
\label{simpd}
\end{eqnarray}

With these simplifications we can now express the luminosity in terms of 
products of symmetric tensors containing only radial components. 
This makes it possible to calculate
these products in closed form, for any multipole order, $l$.
Using the previous expressions and the following rules (see~\ref{appa2}):
\begin{eqnarray}
\SS^l_0\cdot\left[\DD^{(j)}\circ\SS^{l-2j}_0\right]=1,\nonumber\\
\left[{\bf\hat L}\circ\SS^{l-1}_0\right] \cdot
\left[\DD^{(j)}\circ\left[{\bf\hat L}\circ\SS^{l-1-2j}_0\right]\right]=
     \frac{1}{l};
\end{eqnarray}
the products~\eref{prod}, with $T=l-2j$, $2p=2l+2$ and
$v=\Omega r$ are
\begin{eqnarray}
\baril{(l+1)}\cdot\baril{(l+1)}
   &\equiv 2^{-(l-1)} \mu^2v^{2l}\Omega^2 f^2_l(\eta)
     \,{\cal C}_l \,,
\nonumber\\
\barjl{(l+1)}\cdot\barjl{(l+1)}
   &\equiv 2^{-(l-2)}l^{-2}\mu^2v^{2l+2}\Omega^2 
    f^2_{l+1}(\eta)\,{\cal D}_l \,,
\label{mult}
\end{eqnarray}
where ${\cal C}_l$ and ${\cal D}_l$ are given by
\numparts
\begin{eqnarray}
\fl \quad {\cal C}_l =
     \sum_{j=0}^{\left\lf\frac{l}{2}\right\rf}
     \frac{(-1)^j\,2^{2j}\,{l\choose j}\,{l\choose{2j}}}{{{2l}\choose{2j}}}
     \sum_{k=0}^{\lf\frac{l-1}{2}\rf-j}{{l-2j}\choose k}\,
     (l-2j-2k)^{2l+2} \,, \label{calC}\\
\fl \quad {\cal D}_l =
     \sum_{j=0}^{\left\lf\frac{l-1}{2}\right\rf}
     \frac{(-1)^j\,2^{2j}\,{l\choose j}\,{l\choose{2j}}}{{{2l}\choose{2j}}}
     \,(l-2j)\,
     \sum_{k=0}^{\lf\frac{l}{2}\rf-j-1}{{l-1-2j}\choose k}\,
     (l-1-2j-2k)^{2l+2} \,. \label{calD}
\end{eqnarray}
\endnumparts
Finally, the luminosity in terms of ${\cal C}_l$, ${\cal D}_l$, and
$f_l(\eta)$ is
\begin{equation}
\dot E= \dot E_Q\,\left\{1+ \sum_{l=3}^{\infty}  B_l
           f^2_{l}(\eta) v^{2l-4}
         \right\} \,,
\label{lumifinal}
\end{equation}
where $\dot E_Q=\frac{32}{5}\mu^2v^4\Omega^2$ 
is the luminosity due to the quadrupole term and 
$B_l$ is given by
\begin{equation}
B_l =
\frac{5}{16}\,\frac{(l+1)\,(l+2)} {l\,(l-1)\,(2l+1)!}
         \left[{\cal C}_l+\frac{16\,(2l+1)\,l}{(l-2)\,(l+2)}\,{\cal D}_{l-1}
         \right] \,.
\label{bsubl}
\end{equation}

To study the convergence of this series, we apply the standard Cauchy
ratio
test, requiring that
\begin{eqnarray}
\lim_{l \to \infty} {{B_{l+1} f_{l+1}^2(\eta)} \over{B_{l}
f_{l}^2(\eta)}} v^2 < 1 \,.
\end{eqnarray}
Evaluating this numerically up to $l=250$, we find the constraint 
$|f_{l+1}(\eta)/f_l (\eta)|\,v \to (1+\sqrt{1-4\eta})v/2 < 0.74$.  
The case $\eta=0.25$ must be treated separately, since $f_l(0.25)=0$
for odd $l$.  For this case, the Cauchy ratio test takes the form
$|f_{l+2}/f_l|^{1/2}\,v \to v/2 < 0.74$.  which is the continuous
limit of the previous statement.
Note that, for a given $v$, the convergence is worst for $\eta =0$
(test mass limit) and best for $\eta=1/4$ (equal masses).

\section{Black-hole perturbation theory and the bare-multipole
truncation
}

In this section we consider the calculation of the
bare multipole moments in the restricted context of
a binary system with small mass ratio. We shall
therefore demand $\eta \ll 1$
throughout this section. The method of calculation
used here is completely different from the one used 
in the previous two sections.

For the problem considered in this paper, the internal 
structure of the orbiting masses is of no consequence. 
For simplicity, in this section we take the larger mass
$m_1$ to be a nonrotating black hole. The smaller
mass $m_2$ then creates a small perturbation in the
gravitational field of the black hole. This
perturbation propagates away from the source as a
gravitational wave. By virtue of the small mass approximation, the
gravitational perturbations are accurately described
by solving a linear wave equation on the Schwarzschild 
spacetime. This equation is called the Teukolsky equation 
\cite{teukolsky}, and it can be solved exactly, for example, 
using numerical methods \cite{tag}.

We assume that the mass $m_2$ moves on a circular orbit 
with radius $r$, such that $v\equiv (m/r)^{1/2}$ is much 
smaller than unity. (Here, $m=m_1+m_2\simeq m_1$ is the 
total mass.) In this limit the Teukolsky equation can be 
solved {\it analytically}, and the bare multipole moments 
of the radiative field can be evaluated. Such a calculation 
was first carried out by Poisson \cite{poisson}, and then 
completed by Poisson and Sasaki \cite{ps}, 
hereafter referred to as PS.

The Teukolsky equation is analyzed by first separating
the variables. This means that the radiative field is
expressed not in terms of symmetric trace-free tensors, 
but in terms of (spin-weighted) spherical
harmonics \cite{goldberg}. The two representations 
are entirely equivalent \cite{thorne80}, and in both
cases the gravitational-wave luminosity takes the
form of \eref{edotfinal}. Here, of course,
$\eta \to 0$ so that $|f_l(\eta)| = 1$. 

The bare multipole moments were calculated explicitly
by PS, who express the gravitational-wave luminosity 
as
\begin{equation}
\dot{E} = \dot{E}_Q \biggl( 1 + \sum_{l=3}^\infty
\sum_{m=1}^l \eta_{l m} \biggr),
\label{e.2}
\end{equation}
apart from a slight change of notation; cf.\ their 
equation~(5.14). The contributions $\eta_{lm}$ take different 
forms according to whether $l+m$ is odd or even: 
\begin{equation}
\eta_{lm} = \left\{
\begin{array}{ll}
p_{lm}\, v^{2l-4} & \qquad \mbox{for $l+m$ even} \\
q_{lm}\, v^{2l-2} & \qquad \mbox{for $l+m$ odd}
\end{array}
\right. .
\label{e.3}
\end{equation}
The coefficients $p_{lm}$ and $q_{lm}$ are explicitly 
given in equations~(5.15) and (5.16) of PS.

Simple manipulations bring \eref{e.2} into
the form of \eref{edotfinal}, which
we write as
\begin{equation}
\dot{E} = \dot{E}_Q \biggl( 1 + \sum_{l=3}^\infty
B_l\, v^{2l-4} \biggr)\,.
\label{e.4}
\end{equation}
We find
\begin{equation}
B_l = p_{ll} + \sum_{n=0}^{N} \bigl( p_{lm}
+ q_{l-1,m} \bigr)\,.
\label{e.5}
\end{equation}
Here, $m$ is to be considered to be a function
of $n$: for odd $l$, $m=2n+1$ and $N=(l-3)/2$;
for even $l$, $m=2n+2$ and $N=(l-4)/2$. This
ensures that the sum is properly restricted
to values of $m$ such that $l+m$ is 
{\it even}. The largest value of $m$ contributing
to the sum is therefore $l-2$. 

Some algebra, using the explicit expressions for 
$p_{lm}$ and $q_{lm}$ provided by PS, 
converts \eref{e.5} into
\begin{equation}
B_l = \frac{5(l+1)(l+2)}{16(2l+1)! (l-1)l} 
 (1 + b_l)l^{2l+2} \,,
\label{e.6}
\end{equation}
where
\begin{equation}
\fl b_l = \frac{l!^2}{(2l)!  l^{2l+2}} 
\sum_{n=0}^N  \frac{ m^{2l+2} (l-m)!\ (l+m)!}{
[(l-m)/2]!^2   
[(l+m)/2]!^2} 
\Biggl[ 1 + \frac{4(2l-1)(2l+1)(l-m)(l+m)}
{(l-2)(l+2)  m^2} \Biggr]\,.
\label{e.7}
\end{equation}
This expression for $B_l$ is entirely equivalent to that
given in \eref{bsubl}.
We have indeed verified that these $B_l$'s are numerically
equal to the ones derived in section~3. However, we have not 
been able to establish their equality {\it algebraically}.   

We now wish to examine the convergence of the sequence 
$B_l v^{2l}$. This analysis will greatly benefit from
the simplicity of our current expression for $B_l$.
We point out first that the sequence $b_l$ converges.
This was established by numerical experiment, which also
reveals that
$b_\infty \simeq 0.01$.
This implies that the behavior of $B_l$ as 
$l \to \infty$ is determined entirely by 
the factor $l^{2l+2}/(2l+1)!$ in \eref{e.6}.
Writing $(2l+1)!$ as $(2l)^2 \Gamma(2l) [1+O(1/l)]$
and using the Stirling approximation \cite{abram}
\begin{equation}
\Gamma(z) = (2\pi)^{1/2} {\rm e}^{-z} z^{z-1/2} 
\bigl[ 1 + O(1/z) \bigr]\,,
\end{equation}
we quickly arrive at the asymptotic form
\begin{equation}
B_l = \frac{5(1+b_\infty)}{64}\, 
\biggl( \frac{l}{\pi} \biggr)^{1/2}
\biggl( \frac{\mbox{e}}{2} \biggr)^{2l}\, 
\Bigl[ 1 + O \bigl( l^{-1} \bigr) \Bigr] \,.
\label{e.10}
\end{equation}

We therefore find that the sequence $B_l v^{2l}$ behaves
asymptotically as $(\mbox{e} v /2)^{2l}$. A direct
application of the Cauchy ratio test then reveals that
the sequence converges provided 
\begin{equation}
v < 2/\mbox{e} \simeq 0.7358.
\label{e.11}
\end{equation}
The generalization of this result, appropriate for the case 
of nonvanishing mass ratios, was given in section~1.

\section{Concluding remarks}

We have shown that a truncated model for gravitational 
radiation from binary systems in circular orbits converges 
for values of the orbital velocity that encompass
all inspirals  of physical interest.  
However, our model is
admittedly non-physical, and may only be revealing that, whatever
poor convergence properties have been seen to date in the PN
expansion, they do not arise from summing over the Newtonian 
part of the multipole moments, but arise instead 
from the PN corrections to the moments.  

Figure 4 illustrates the limitations of our
truncation and reinforces the notion that it is unphysical.  Shown is
a comparison, in the test body limit, between our bare-multipole
series and the physical results of black-hole perturbation theory, 
including both the numerical ``exact'' results \cite{eric2} and 
the true PN series, accurate to 3.5PN order \cite{tagsasak}. It is 
apparent that the truncated model compares rather poorly; this is 
mostly due to the fact that while our series is a sum of positive
terms, the true PN series is alternating. 
Nevertheless, our analysis can be improved somewhat.  
The lowest-order tail corrections to each STF moment can be written
down explicitly, and could thus be added in a straightforward way.  
It is possible that the $O(\epsilon)$ corrections to each moment 
could also be calculated without too much difficulty.  
Whether such a ``dressed-multipole'' truncation shows better agreement
with the exact results --- and still converges --- will be a subject for
future work.

\appendix

\section{Symmetric products of unit vectors $\SS^{l-k}_k$\label{appa}}

The multipole moments in~\eref{2pnmults}
can be expressed in terms of products of components of unit vectors, 
\begin{equation}
\left(\PP_k^{l-k}\right)^{i_1\cdots i_l}\equiv
\hat\la^{i_1} \hat\la^{i_2}\cdots\hat\la^{i_k} 
  \hat n^{i_{k+1}} \hat n^{i_{k+2}}\cdots \hat n^{i_l}.
\label{p-def}
\end{equation}
Here $k$ components are in the tangential direction to the circular orbit,
$\hat \la^i$, and $(l-k)$ are in the radial direction, $\hat n^i$,
with ${\bf\hat n}\cdot{\bf\hat n}= \hat n^i \hat n^i=1$,
$\hat{\boldla}\cdot\hat{\boldla}=\hat\la^i\hat\la^i=1$, 
${\bf\hat n}\cdot\hat{\boldla}=\hat n^i \hat \la^i=0$,
and ${\bf\hat n}\times\hat{\boldla}={\bf\hat L}$ 
(or $\hat L^i=\levi_{ijk} \hat n^j \hat \la^k$).

A symmetric tensor can be constructed from 
$\left(\PP_k^{l-k}\right)^{i_1\cdots i_l}$, 
\begin{eqnarray}
\left(\SS_k^{l-k}\right)^{i_1\cdots i_l} &\equiv
\hat\la^{(i_1} \hat\la^{i_2}\cdots\hat\la^{i_k} 
  \hat n^{i_{k+1}} \hat n^{i_{k+2}}\cdots \hat n^{i_l)}   \nonumber\\
 &\equiv\frac{1}{\cal N}\sum_{\Pi_{\cal N}}
	  \left[\hat\la^{i_1} \hat\la^{i_2}\cdots\hat\la^{i_k} 
                \hat n^{i_{k+1}} \hat n^{i_{k+2}}\cdots \hat n^{i_l} \right]  
\nonumber\\
 &\equiv \frac{1}{\cal N}\sum_{\Pi_{\cal N}} 
\left(\PP_k^{l-k}\right)^{i_1\cdots i_l}\,,
\label{s-def}
\end{eqnarray}
where the sum is over all the ${\cal N}\equiv{l\choose k}$ 
permutations of the indices $\{i_l\}$.

For simplicity, from now on we will drop the set of indices 
$\{i_1\cdots i_l\}$ in the notation for $\PP_k^{l-k}$ and $\SS_k^{l-k}$.

\subsection{Differentiation rule for $\SS^{l-k}_k$\label{appa1}}

In a system of two bodies following a circular orbit, the time derivative 
of the components of the normal and tangential unit vectors are, 
respectively,
\begin{equation}
\frac{d \hat n^i}{dt}=\Omega \hat \la^i \,,\qquad 
\frac{d \hat \la^i}{dt}=-\Omega \hat n^i,
\end{equation}
where $\Omega$ is the angular velocity, such that $v=\Omega r$.

We need to calculate the time derivative of a symmetrized product of $l$
components of unit vectors, where $k$ of them are tangential ($\hat \la^i$'s) 
and $(l-k)$ are normal ($\hat n^i$'s), as in~\eref{s-def}.
The time derivative of one of the $\PP_k^{l-k}$'s that 
belongs to $\SS_k^{l-k}$:
\begin{eqnarray}
{d \over {dt}}\left(\PP_k^{l-k}\right) =\Omega\Big[
&-\Big(\hat n^{i_1}\,\hat\la^{i_2}\cdots\hat\la^{i_k}\,\,\,\hat n^{i_{k+1}}\hat n^{i_{k+2}}
\cdots \hat n^{i_l} \nonumber\\
&+\hat\la^{i_1}\,\hat n^{i_2}\cdots\hat\la^{i_k}\,\,\,\hat n^{i_{k+1}}\hat n^{i_{k+2}}\cdots \hat n^{i_l} 
\nonumber\\
&\qquad\qquad \cdots\cdots\cdots  \nonumber\\
&+\hat\la^{i_1}\hat\la^{i_2}\cdots \hat n^{i_k}\,\,\,\hat n^{i_{k+1}}\hat n^{i_{k+2}}\cdots \hat n^{i_l}
\Big) \nonumber\\
&+\Big(\hat\la^{i_1}\hat\la^{i_2}\cdots\hat\la^{i_k}\,\,\,\hat\la^{i_{k+1}}\,\hat n^{i_{k+2}}
\cdots \hat n^{i_l} \nonumber\\
&+\hat\la^{i_1}\hat\la^{i_2}\cdots\hat\la^{i_k}\,\,\,\hat n^{i_{k+1}}\,\hat\la^{i_{k+2}}
\cdots \hat n^{i_l} \nonumber\\
&\qquad\qquad \cdots\cdots\cdots  \nonumber\\
&+\hat\la^{i_1}\hat\la^{i_2}\cdots\hat\la^{i_k}\,\,\,\hat n^{i_{k+1}}\hat n^{i_{k+2}}\cdots\hat\la^{i_l} 
\Big) \Big]\, ,
\label{p-deriv}
\end{eqnarray}
is the sum of two parts: the first parenthesis with $k$ terms of the
type $\PP_{k-1}^{l-k+1}$, and the second parenthesis with $(l-k)$ terms
of the type $\PP_{k+1}^{l-k-1}$.

Taking the derivative of the whole $\SS_k^{l-k}$ creates ${l\choose k}$
sets of terms like~(\ref{p-deriv}). In the expression for 
$d\left(\SS_k^{l-k}\right)/dt$, all permutations of indices will be present 
because they were present in the original $\SS_k^{l-k}$. We just need to
collect all the terms of the type $\PP_{k-1}^{l-k+1}$, that belong to 
$\SS_{k-1}^{l-k+1}$, and all the terms of the type $\PP_{k+1}^{l-k-1}$, 
that belong to $\SS_{k+1}^{l-k-1}$.
There are ${l\choose{k-1}}$ $\PP_{k-1}^{l-k+1}$-terms
and ${l\choose{k+1}}$ $\PP_{k+1}^{l-k-1}$-terms. So the expression for 
the time derivative of $\SS_k^{l-k}$ is
\begin{eqnarray}
{d \over {dt}} \left(\SS_k^{l-k}\right)&=\frac{1}{{l\choose k}}\,\Omega\left[
 \frac{{l\choose k}\,(l-k)}{{l\choose{k+1}}}
\sum_{\Pi_{j}\{i_j\}} \PP_{k+1}^{l-k-1}
 -\frac{{l\choose k}\,k}{{l\choose{k-1}}}
\sum_{\Pi_{j}\{i_j\}} \PP_{k-1}^{l-k+1}\right]  \nonumber\\
&= \Omega\left[(l-k)\,\SS_{k+1}^{l-k-1}-k\,\SS_{k-1}^{l-k+1}\right].
\label{s-deriv}
\end{eqnarray}
For this procedure to be valid, at least one of each of the new symmetrized 
products $\SS_{k+1}^{l-k-1}$ and $\SS_{k-1}^{l-k+1}$ has to be obtained.
In other words, 
\begin{equation}
 \frac{{l\choose k}\,(l-k)}{{l\choose{k+1}}}\ge 1,\quad\mbox{\rm and}\quad
 \frac{{l\choose k}\,k}{{l\choose{k-1}}}\ge 1 \,,
\end{equation}
which are trivially true, with the equality corresponding to $k=l$ (for
$\SS^0_l$), and $k=0$ (for $\SS^l_0$), the two possible limit cases.
Then,~\eref{s-deriv} is the rule of differentiation for the symmetric
tensors $\SS_k^{l-k}$.

\subsection{Products of $\SS^T_0$'s\label{appa2}}

In this section we will limit the discussion to symmetric products
that contain only radial components. In this case, (with the
same conventions as in~(\ref{symmet}), (\ref{delta-op}), 
and~(\ref{delta-num})), we have
\begin{eqnarray}
\fl \SS^T_0= \hat n^{(i_1} \hat n^{i_2}\cdots \hat n^{i_T )}
       = \hat n^{i_1} \hat n^{i_2}\cdots \hat n^{i_T}\,,\\
\fl {\bf\hat L}\circ\SS^{T-1}_0=\hat L^{(i_1} \hat n^{i_2}\cdots 
	\hat n^{i_T)}
       = \frac{1}{T}\sum_{\Pi_T}
       \hat L^{i_1} \hat n^{i_2}\cdots \hat n^{i_T}\,,\\
\fl \DD^{(j)}\circ\SS^{T-2j}_0=
    \delta^{(i_1 i_2} \delta^{i_3 i_4}\cdots\delta^{i_{2j-1},i_{2j}} 
    \hat n^{i_{2j+1}}\cdots \hat n^{i_T )}\nonumber\\
    =\frac{1} {{\cal N}(T;j,2)} \sum_{\Pi}
    \delta^{i_1 i_2} \delta^{i_3 i_4}\cdots\delta^{i_{2j-1},i_{2j}} 
    \hat n^{i_{2j+1}}\cdots \hat n^{i_T}\,,\\
\fl \DD^{(j)}\circ\left[{\bf\hat L}\circ\SS^{T-1-2j}_0\right]=
    \delta^{(i_1 i_2} \delta^{i_3 i_4}\cdots\delta^{i_{2j-1},i_{2j}} 
    \hat L^{i_{2j+1}} \hat n^{i_{2j+2}}\cdots \hat n^{i_T )}\nonumber\\
    =\frac{1} {(T-2j) {\cal N}(T-1;j,2)} \sum_{\Pi}
    \delta^{i_1 i_2} \cdots\delta^{i_{2j-1},i_{2j}} 
    \hat L^{i_{2j+1}} \hat n^{i_{2j+2}}\cdots \hat n^{i_T}\,,
\end{eqnarray}
where $\Pi$ in the latter two equations simply denotes the distinct
permutations, and 
\begin{equation}
{\cal N}(T;j,2) \equiv \frac{1}{j!}\,{T\choose 2}\,
                {{T-2}\choose 2}\,\cdots\,{T-2j\choose 2}\,.
\end{equation}
Using the orthonormality of ${\bf\hat n}$ and ${\bf\hat L}$, and the
definition
$\DD^{(1)}\cdot {\bf\hat n}{\bf\hat n}=
\delta^{ij}\,\hat n^i \hat n^j=1$, 
we calculate the products:
$\left[\DD^{(j)}\circ\SS^{T-2j}\right]\cdot\SS^T_0$ and
$\left[\DD^{(j)}\circ\left[{\bf\hat L}\circ\SS^{T-1-2j}_0\right]\right]
\cdot \left[{\bf\hat L}\circ\SS^{T-1}_0\right]$, which occur in 
\eref{prod}.
Notice that we use as ``target" (or second factor in the product), the
simpler of the two factors. This simplifies the counting of resulting
terms. For both cases, the expression is, schematically,
\begin{equation}
\left[\frac{1}{{\cal N}_1}\sum \mbox{\rm Term}_1\right]\cdot
\left[\frac{1}{{\cal N}_2}\sum \mbox{\rm Term}_2\right]\,,
\end{equation}
where ${\cal N}_i$ is the number of distinct terms in each sum.
Next, we calculate the effect of multiplying one term only
from the symmetric first factor by the whole of the target, say
\begin{equation}
\left[\mbox{\rm Term}_1\right]\cdot
\left[\frac{1}{{\cal N}_2}\sum \mbox{\rm Term}_2\right].
\label{one-term}
\end{equation}
There are ${\cal N}_1$ identical products like this in the final
result, so
\begin{equation}
\left[\frac{1}{{\cal N}_1}\sum \mbox{\rm Term}_1\right]\cdot
\left[\frac{1}{{\cal N}_2}\sum \mbox{\rm Term}_2\right]=
\left[\mbox{\rm Term}_1\right]\cdot
\left[\frac{1}{{\cal N}_2}\sum \mbox{\rm Term}_2\right].
\end{equation}
For the first product, ~\eref{one-term} gives
\begin{equation}
\left[\delta^{i_1 i_2} \delta^{i_3 i_4}\cdots\delta^{i_{2j-1},i_{2j}}
    \hat n^{i_{2j+1}}\cdots \hat n^{i_T}\right]\cdot
   \left[\hat n^{i_1} \hat n^{i_2}\cdots \hat n^{i_T}\right]=1 \,,
\end{equation}
trivially, as the target has only one term. The result is
\begin{equation}
\left[\DD^{(j)}\circ\SS^{T-2j}\right]\cdot\SS^T_0=1.
\end{equation}
For the second product,~\eref{one-term} gives
\begin{equation}
\fl \left[\delta^{i_1 i_2} \delta^{i_3 i_4}\cdots\delta^{i_{2j-1},i_{2j}}
  \hat L^{i_{2j+1}} \hat n^{i_{2j+2}}\cdots \hat n^{i_T}\right]\cdot
  \left[\frac{1}{T}\sum_{\Pi_T} \hat L^{i_1} \hat n^{i_2}\cdots \hat
n^{i_T}\right]=
\frac{1}{T}\,,
\end{equation}
because only one of the $T$ terms in the target has the angular
momentum component in position $i_{2j+1}$, as in the first factor. The
result is
\begin{equation}
\left[\DD^{(j)}\circ\left[{\bf\hat L}\circ\SS^{T-1-2j}_0\right]\right]
\cdot \left[{\bf\hat L}\circ\SS^{T-1}_0\right]=\frac{1}{T}\,.
\end{equation}

\subsection{Canonical form of an even number of derivatives of 
$\SS^T_0$\label{appc}}

Here we discuss the derivation of 
an expression for the $(2p)$-derivative of
$\SS^T_0$ in which all terms are of rank $T$.  We will carry out the
derivation
explicitly only for the terms proportional to $\SS^T_0$, since only
these survive the contraction in \eref{prod}.
We can illustrate the problem appearing in the discussion of~\Sref{mysect}
following
\eref{der-sinecosine}, 
with this example: take, say, $T=4$ and $p=2$ 
in~(\ref{der-symmeta}). In the sine--cosine representation, this is
\begin{eqnarray}
{d^4 \over {dx^4}}(\sin^4\Omega x)=&\Omega^4\Bigl[\left(32\,\sin^4\Omega x
   -192\,\sin^2\Omega x\,\cos^2\Omega x +32\,\cos^4\Omega x\right) \nonumber\\
&+\left(8\sin^2\Omega x -8\cos^2\Omega x\right)\Bigr]\,.
\label{oops1}
\end{eqnarray}
The first parenthesis contains 
products of ``rank''~4 while the second contains 
products of ``rank''~2. Direct differentiation, however, gives
\begin{equation}
{d^4 \over {dx^4}} (\sin^4\Omega x)=\Omega^4\bigl[40\,\sin^4\Omega x
-192\,\sin^2\Omega x\,\cos^2\Omega x +24\,\cos^4\Omega x \bigr]\,,
\label{oops2}
\end{equation}
where all the terms are of ``rank''~4. The expansion of $\sin^T\Omega x$
in terms of sines and cosines of multiples angles before differentiation
produced this effect. However, equations~(\ref{oops1}) and~(\ref{oops2})
are identical: in~(\ref{oops1}), multiplying the second parenthesis by
$\cos^2 \Omega x + \sin^2\Omega x =1$ yields 
$8\sin^4 \Omega x - 8\cos^4\Omega x $. 

We need to perform the same regrouping of terms 
on~(\ref{der-symmeta}, \ref{der-symmetb}). This is very difficult to do in
generic form. However, we should remember that these derivatives are to be
multiplied by a factor proportional to $\SS^l_0$, which has components
only 
in the radial direction.  Consequently,
all terms in the derivatives possessing tangential 
components (terms with cosines in the sine--cosine representation) will have
no effect.
The only surviving term in~(\ref{der-symmeta}, \ref{der-symmetb})
is the one that has only radial components (all-sines term).
Explicitly, this
condition is $T-2k-2q=0$ in~(\ref{der-symmeta}) and 
$T-2k-2q-1=0$ in~(\ref{der-symmetb}).

One can demonstrate that in the regrouping process, the coefficient of 
the final all-sines term is obtained by simply
adding the coefficients of all the
all-sines terms of different rank $T-2k$ in the original sum (the
same conclusion holds for the all-cosines term). 
The demonstration
generalizes the example above,
and involves
multiplying the lowest ranked terms by $\cos^2 \Omega x + \sin^2\Omega x$,
combining with the next higher ranked terms, and repeating the process
until all terms are the same rank.  

In ~(\ref{der-symmeta}, \ref{der-symmetb}), we thus need to add the 
coefficients of all $\SS^{2q}_{T-2k-2q}$, for $T-2k-2q=0$ and all $k$, 
in~(\ref{der-symmeta}), and the coefficients of all $\SS^{2q+1}_{T-2k-2q-1}$, 
for $T-2k-2q-1=0$ and all $k$, in~(\ref{der-symmetb}), to get the coefficient
of $\SS^T_0$.

For~(\ref{der-symmeta}), the condition $2q=T-2k$, eliminates all the
terms in the sum over $q$ except the last one, so the coefficient of
$\SS^{T-2k}_0$ is
\begin{equation}
\fl(-1)^{\left\lf\frac{T}{2}\right\rf -k}{T\choose k}\,(T-2k)^{2p}\,
(-1)^{\left\lf\frac{T}{2}\right\rf -k}\,{{T-2k}\choose{T-2k}}=
{T\choose k}\,(T-2k)^{2p}.
\end{equation}
For~(\ref{der-symmetb}), the condition $2q+1=T-2k$, also eliminates all the
terms in the sum over $q$ except the last one, so the coefficient of
$\SS^{T-2k}_0$ is
\begin{equation}
\fl(-1)^{\left\lf\frac{T-1}{2}\right\rf +k}{T\choose k}\,(T-2k)^{2p}\,
(-1)^{\left\lf\frac{T-1}{2}\right\rf -k}\,{{T-2k}\choose{T-2k}}=
{T\choose k}\,(T-2k)^{2p}.
\end{equation}
The result is
the same for $T$ even or odd.  Thus, the term in the $(2p)$ 
derivative of $\SS^T_0$ that will survive the product by $\SS^l_0$ is:
\begin{equation}
\fl\left.{}^{(2p)}\SS^T_0\right|_{\mbox{\rm all-sines}}=
\left[\frac{(-1)^p\,\Omega^{2p}}{2^{T-1}}
\sum_{k=0}^{\left\lf\frac{T-1}{2}\right\rf}
{T\choose k}\,(T-2k)^{2p}\right]\SS^T_0
\equiv {\cal A}(T,p)\,\SS^T_0.
\label{c-result}
\end{equation}

\section{Mass functions $f_l(\eta)$\label{appb}}
In a general coordinate system, the expressions for
the leading order of the radiative multipole moments are simply
\begin{eqnarray}
\bari^{\brl}= \left\{
   m_1  x_1^{i_1}x_1^{i_2}\cdots x_1^{i_l}
+  m_2  x_2^{i_1}x_2^{i_2}\cdots x_2^{i_l}
\right\}_{{\rm STF}}, \nonumber\\
\barj^{\brl}=\left\{\levi_{i_1 a b}\left[
   m_1  x_1^a v_1^b   x_1^{i_2}\cdots x_1^{i_l}
+  m_2  x_2^a v_2^b   x_2^{i_2}\cdots x_2^{i_l}\right]
\right\}_{{\rm STF}}.
\end{eqnarray}
Changing to the CM-coordinate system defined by
the Newtonian relations (consistent with our bare-multipole
approach)
$m_1{\bf x}_1 +m_2{\bf x}_2=0$, 
and with ${\bf x} \equiv{\bf x}_1-{\bf x}_2$, and
\begin{eqnarray}
{\bf x}_1=\frac{m_2}{m} {\bf x}\,,\qquad {\bf x}_2=-\frac{m_1}{m}\,{\bf x}\,,\\
{\bf v}_1=\frac{m_2}{m} {\bf v}\,,\qquad {\bf v}_2=-\frac{m_1}{m}\,{\bf v}\,,
\end{eqnarray}
the moments become 
\begin{eqnarray}
\barix{l}= \left[ 
  m_1\left(\frac{m_2}{m}\right)^l 
 +(-1)^l\, m_2\left(\frac{m_1}{m}\right)^l \right]
{\bf x}^{\brl}_{{\rm STF}},
\nonumber\\
\barjx{l}= \left[
  m_1\left(\frac{m_2}{m}\right)^{l+1} 
 +(-1)^{l+1}\, m_2\left(\frac{m_1}{m}\right)^{l+1} \right] 
\mu^{-1} \left[{\bf L}\circ {\bf x}^{{<}l-1{>}} \right]_{{\rm STF}},
\label{cmb}
\end{eqnarray}
where $\bf L =\mu {\bf x} \times {\bf v}$ is the angular momentum vector, and
${\bf x}^{\brl}\equiv x^{i_1}x^{i_2}\dots x^{i_l}$.
We define $\rho$ such that
\begin{equation}
\fl m_2=\rho\, m_1\,,\quad m=(1+\rho)\,m_1\,,
\quad \mu=\frac{\rho}{(1+\rho)^2}\,m\,,
\quad \eta=\frac{\rho}{(1+\rho)^2}\,,
\end{equation}
with the convention $m_2 \le m_1$.
Inverting the last expression, we have
\begin{equation}
\rho(\eta)=\frac{1}{2\eta}\,\left[1-2\eta - \sqrt{1-4\eta}\right].
\label{rho}
\end{equation}
In terms of
$\mu$ and $\rho$, \eref{cmb} becomes
\begin{eqnarray}
\fl\barix{l}=
  \mu\,\frac{\left[\rho^{l-1}+(-1)^l\right]}{(1+\rho)^{l-1}}\,
{\bf x}^{\brl}_{{\rm STF}}
\equiv \mu\,f_{l}(\eta)\,{\bf x}^{\brl}_{{\rm STF}},
\nonumber\\
\fl\barjx{l}= 
  \frac{\left[\rho^l+(-1)^{l+1}\right]}{(1+\rho)^l}\,
  \left[{\bf L}\circ {\bf x}^{{<}l-1{>}} \right]_{{\rm STF}}
\equiv f_{l+1}(\eta) \left[{\bf L}\circ {\bf x}^{{<}l-1{>}} 
        \right]_{{\rm STF}},
\label{fnb}
\end{eqnarray}
Notice that we have not used any information about the trajectory of 
the bodies in this section. This expression for $f_l(\eta)$ is then 
valid for any system of two bodies.

Figure 3 shows the dependence of $f_l$ on the reduced mass $\eta$,
for odd and even values of $l$. Limiting values for $f_l$ in the range
$\eta=(0$, $0.25]$, are
$|f_l(0)|= 1$,
$f_l(0.25)=0$, if $l$ is odd, and 
$f_l(0.25)=1/2^{l-2}$, if $l$ is even.

\ack

This research is supported in part by the 
National Science Foundation
Grants No. PHY 92-22902 and 96-00049 and
by the National
Aeronautics and Space Aministration Grant No. NAGW 3874 at Washington
University.   The work of SWL and EP was supported 
by the Natural Sciences and Engineering Research
Council of Canada.  We are grateful to Matt Visser for helpful
discussions.

\Figures

\begin{figure}
\caption{STF multipole moments displayed in an array with multipole
order increasing to the right, and the order of post-Newtonian corrections
of each multipole moment increasing downward.  Tail terms first appear at
$O(\epsilon^{3/2})$ for each moment.  For each $l$, $\barix{l}$ and
$\barjx{l-1}$ are grouped together.  The polygons below indicate
the moments and PN accuracies required to calculate the indicated
quantity through 2PN order.}
\end{figure}

\begin{figure}
\caption{Fractional difference between series truncated at the
labelled PN order and the exact series, as a function of $v$, for  
$\eta=0$ (solid lines) and $\eta=0.25$ (dashed lines).  Innermost
stable orbit for test body motion is at $v \approx 0.4$.
In the equal-mass case, vanishing of odd-parity moments leads to
degeneracy between adjacent PN approximations.
}
\end{figure}

\begin{figure}
\caption{ The function $|f_l(\eta)|$, for odd 
(dashed lines) and even (solid lines) 
values of $l$.  Note that  $f_2(\eta) \equiv 1$.}
\end{figure}

\begin{figure}
\caption{A comparison between the bare-multipole series (upper
dashed curve), the ``exact'' numerical results from black-hole
perturbation theory (solid curve), and the true PN series
truncated to 3.5PN order (lower dashed curve). The three curves
are plots of $\dot{E}/\dot{E}_Q$ as a function of orbital
velocity $v$, in the physically relevant interval $0<v<0.4$. 
The comparison is valid in the test-body limit $\eta \ll 1$.}
\end{figure}

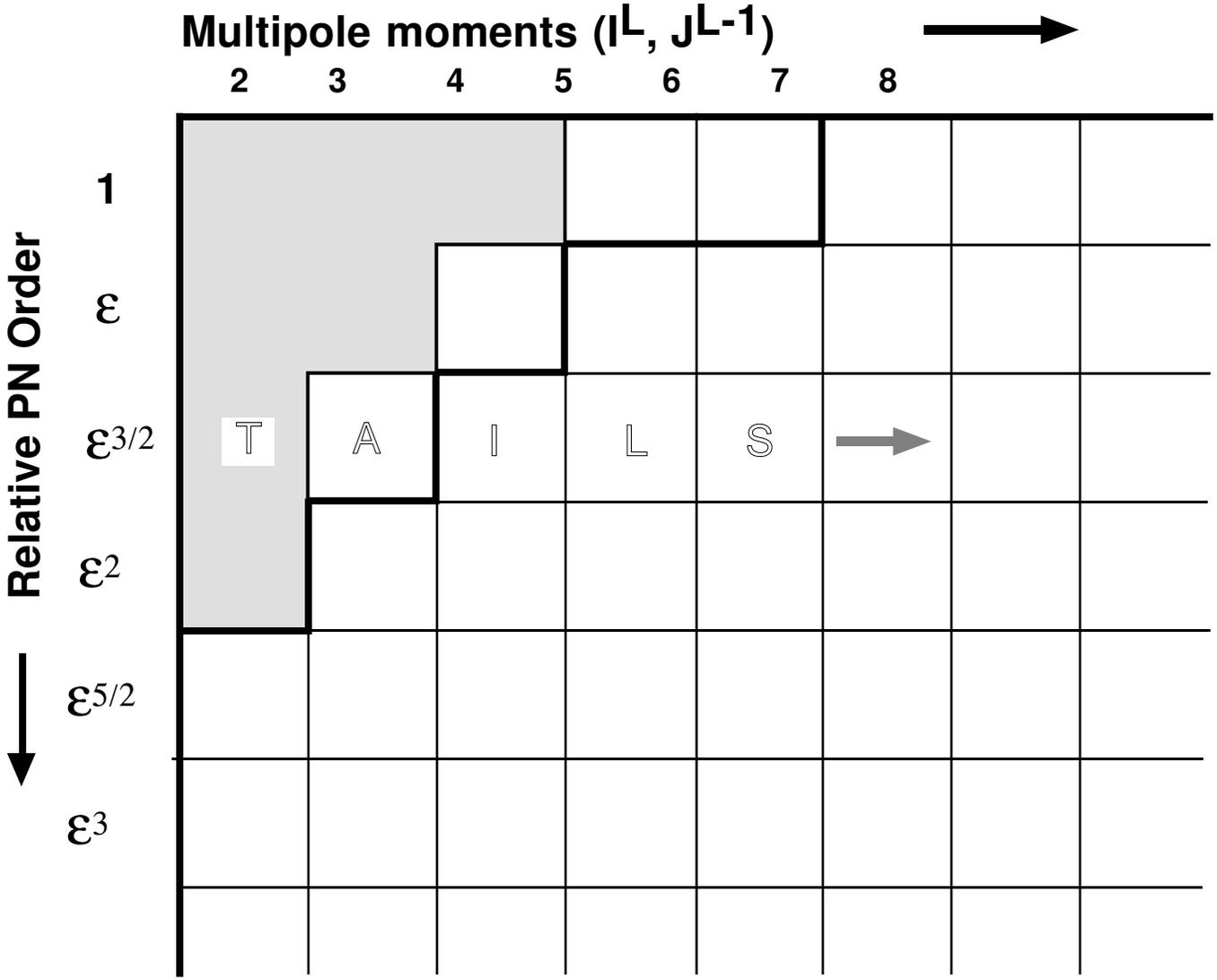
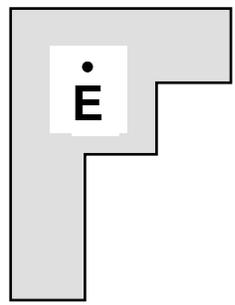
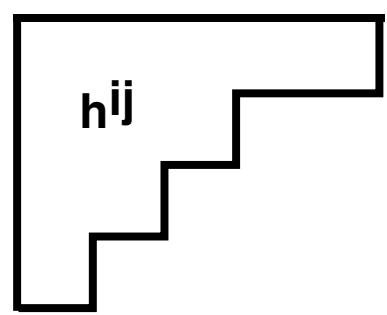
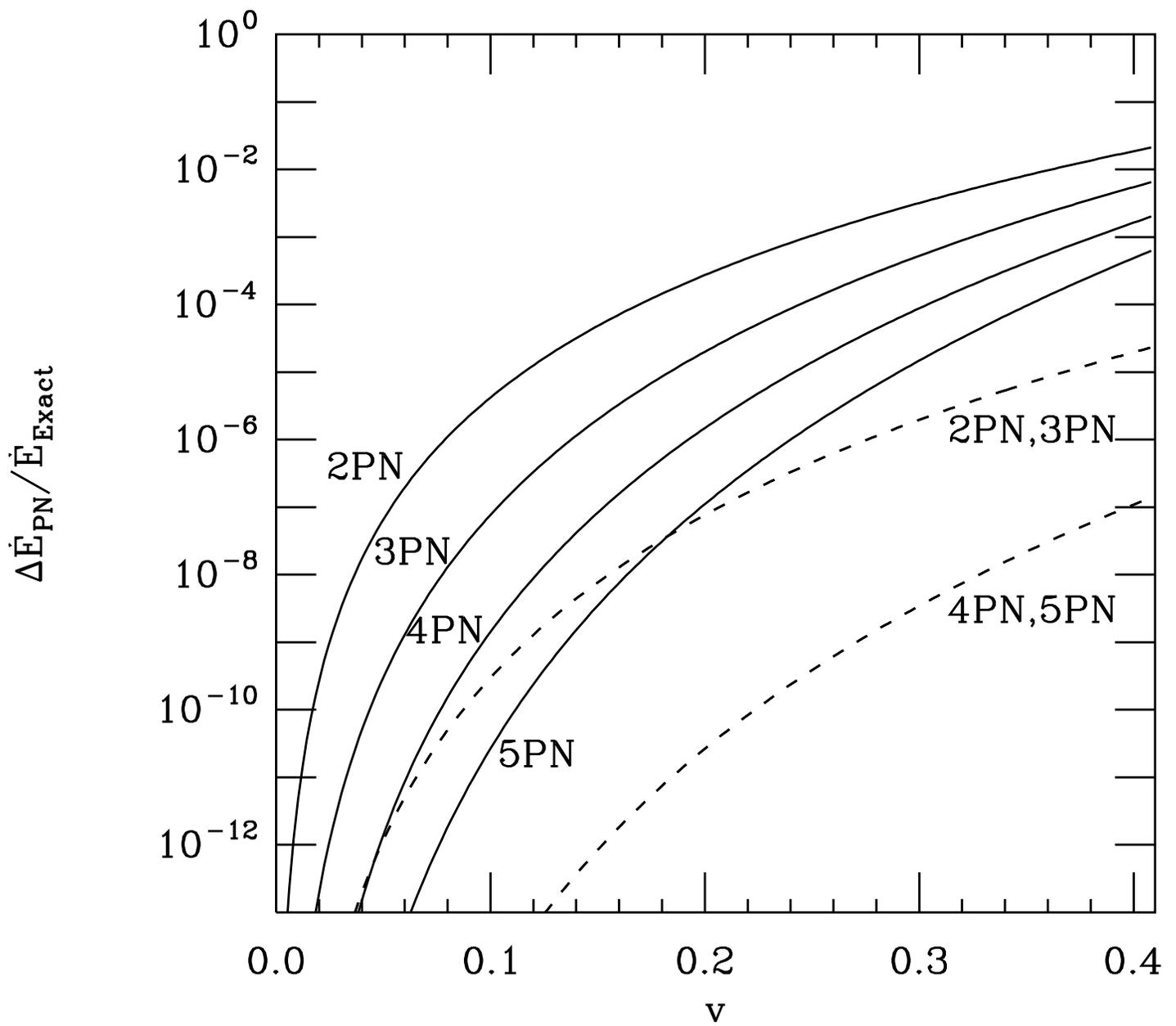
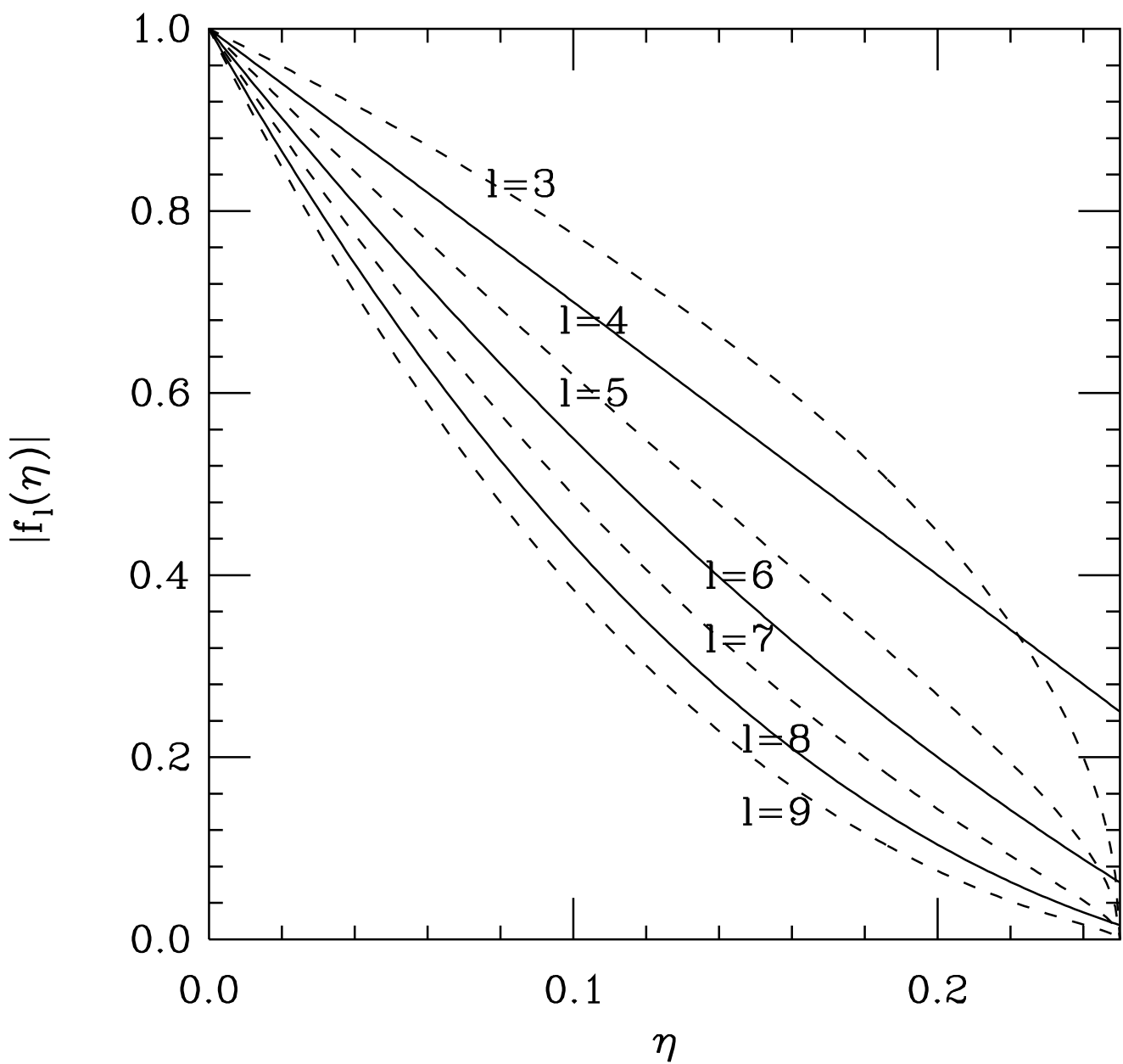
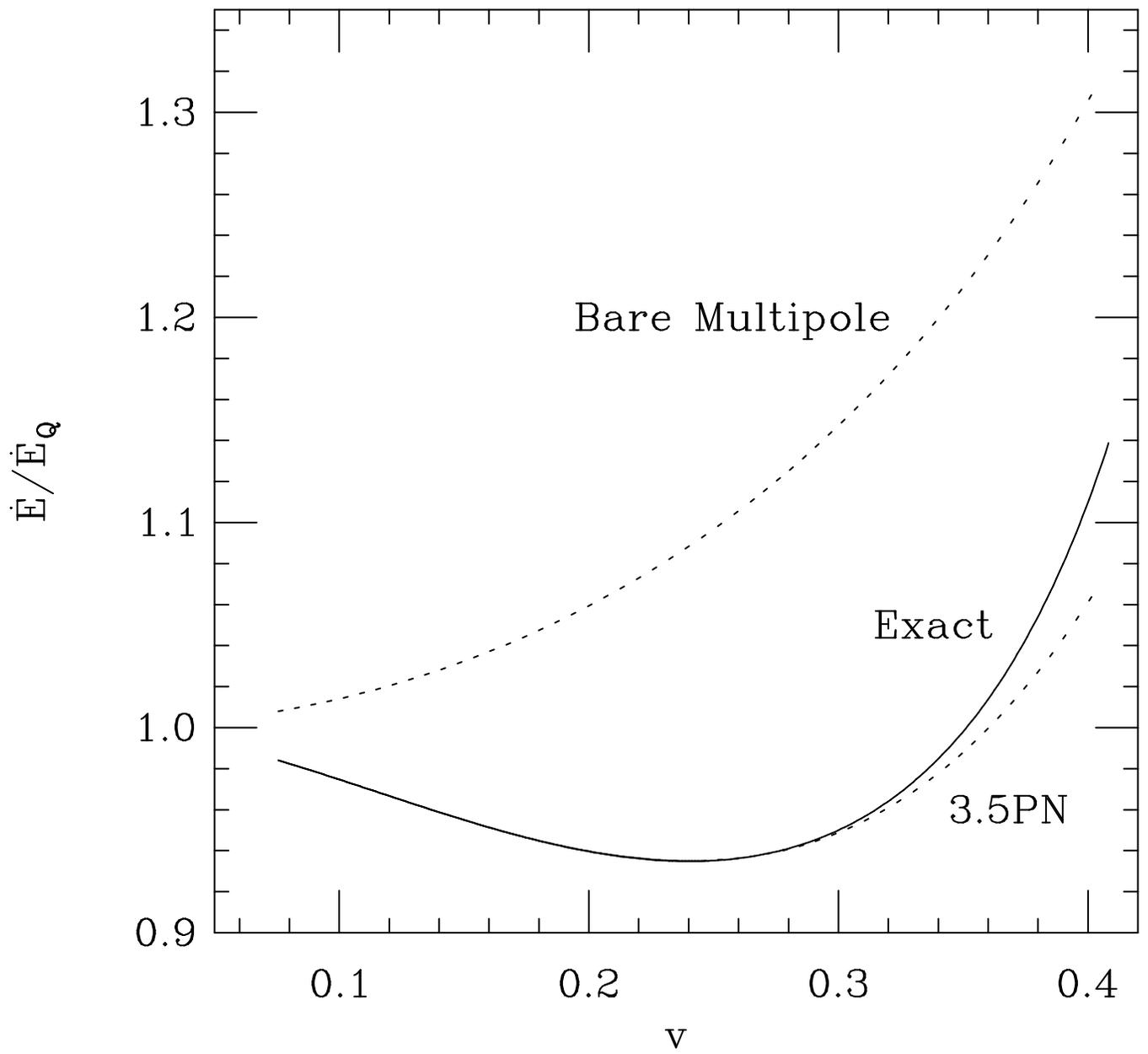
\end{document}